\documentclass[twocolumn,showpacs,preprintnumbers,amsmath,amssymb]{revtex4}
\usepackage{epsfig,amsopn}
\usepackage{graphicx}
\usepackage{epstopdf}
\usepackage{sidecap}
\usepackage{amsmath,amssymb}
\usepackage{amsthm}
\usepackage{enumerate}
\usepackage{bbold}
\newcommand\bea{\begin{eqnarray}}
\newcommand\eea{\end{eqnarray}}
\newcommand\beq{\begin{equation}}  
\newcommand\eeq{\end{equation}}

\newcommand{\bib}{\bibitem}

\newcommand{\non}{\nonumber}  

\newcommand\bs{\boldsymbol}

\newcommand\s{\sigma}
\newcommand{\p}{\tilde{\Psi}}
\newcommand{\ps}{\tilde{\Phi}}

\newcommand{\X}{\tilde{\chi}}

\def\br{{\bs r}}

\begin{document}
\title{Proximity induced superconductivity in Weyl semi-metals}
\author{ Udit Khanna$^{1}$, Arijit Kundu$^{2}$, Saurabh Pradhan$^1$ and Sumathi Rao$^{1,3}$}
\affiliation{  \mbox{$^1$} {Harish-Chandra Research Institute, Chhatnag Road, 
Jhunsi, Allahabad 211 019, India.} \\
\mbox{$^2$}{Department of Physics, Indiana University, 727 East Third Street, Bloomington, IN47405-7105, U.S.A}\\
\mbox{$^3$}{International Center for Theoretical Studies, Tata Institute of Fundamental Research, IISc Campus, Bangalore, 560012, 
India
}}
\begin{abstract}

We study superconducting proximity effects in Weyl semi-metals (WSM) with broken time reversal symmetry by tunnel coupling one of its surfaces to an $s$-wave superconductor using the Green\rq{}s function approach. We find that the band structure  develops coherence peaks, but despite the presence of metallic states in the bulk, the coherence peaks  do not extend far into the bulk and remain confined to a few layers close to the interface, similar to the proximity effect in the topological 
insulators (TI) which are gapped in the bulk. The Weyl nodes remain unaffected, and in that sense, no true gap develops. We also study the induced $p$ and $s$-wave pairing amplitudes classified by their symmetries, as a function of the various parameters of the theory and note the exponential decay of the induced pairings in the bulk both in the TI and the WSM, even at finite chemical potential.
\end{abstract} 
\pacs{74.45.+c, 73.20.At, 74.78.Na}
\maketitle
\section{Introduction} 

In recent years, topology has become an important tool in classifying the phases of matter\cite{reviews}.
Although  the study of topological phases started with the discovery of the quantum Hall
and fractional quantum Hall phases \cite{fqhereviews}
in the eighties, it gained momentum with the discovery of the 
time-reversal invariant topological insulators\cite{TI} a few years ago.  Topological insulators are classified in
terms of their bulk band-structure and by now, there has been a complete classification of
free fermion topological insulators in the presence of disorder\cite{ludwig,kitaev}. All these phases 
have topologically non-trivial momentum space structure, are insulating in the bulk and
have metallic surface states.

It has been generally assumed that it is the gap in the bulk electronic spectrum which makes the 
topologically non-trivial ground state  with its surface states, stable,  and unable to 
decay to the topologically trivial phase.  However, more recently it has been shown that it is possible to have non-trivial momentum space topology even for gapless
fermionic systems. One such recently identified system is the 
Weyl semi-metal phase\cite{vishwanath, balents} which 
has isolated gapless points (weyl nodes) in the bulk spectrum, where exactly 2 bands touch. 
The low energy behaviour close to these points is given by a Weyl hamiltonian of fixed chirality. 
The Weyl nodes are topologically protected, because a gap cannot be opened unless two nodes of opposite chirality are coupled. 
The band structure shows unusual surface states called Fermi arcs\cite{vishwanath}, which
has led to many interesting work\cite{many}. 
The topological
response of the phase has been argued to be a realization of the Adler-Bell-Jackiw anomaly\cite{anomaly}
in condensed matter systems. There are several recent reviews\cite{wsmreviews} which have
discussed various interesting properties of Weyl semimetals.

The introduction of superconductivity  in topological insulators, characterises  a new exotic phase, the topological superconductor,  
whose inherent particle-hole symmetry leads to 
surface states that support Majorana fermions. 
The superconductor doped  TI  $\text{Cu}_x\text{Bi}_2\text{Se}_3$\cite{hor} 
was theoretically predicted\cite{fuberg, sato} to be a TI and experimental evidence was obtained\cite{ando} using
point-contact spectroscopy to detect the itinerant massless Majorana state on the surface.
The introduction of superconductivity via the proximity effect\cite{fukane} 
has also led to considerable work on topological
insulator-superconductor hybrid 
junctions\cite{stanescu,sitthison,tanaka,linder} with special attention to the surface states that develop between them. 
Proximity with an $s$-wave superconductor was shown to lead to a significant renormalization
of the parameters in the effective model for surface states. It was also shown that
when the fermi surface is 
close to the surface Dirac cone vertex, the electrons exhibit $s$-wave pairing, but away from the
vertex, the triplet component increases in amplitude. A full symmetry classification
of all the induced pairings for proximity to  $s$-wave, $p$-wave and $d$-wave superconductors
was  also studied\cite{balatsky,balatsky2} and it was shown that the different induced pairing amplitudes
modify the density of states at the interface significantly\cite{tanakarev,balatskyrev}.

\begin{figure}
 \includegraphics[width=0.45\textwidth]{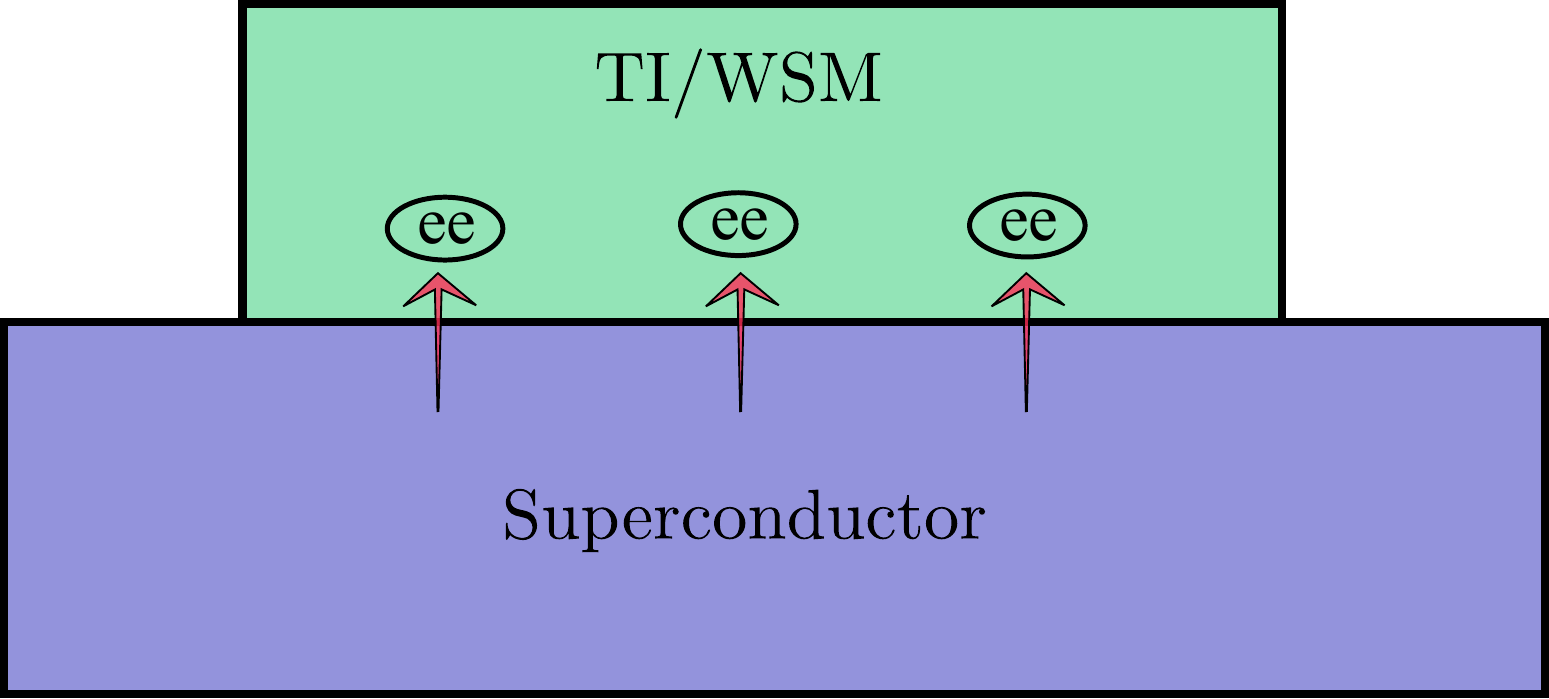}
\caption{(Color online) Schematic of our setup - an $s$-wave superconductor is coupled with the Weyl semi-metal/topological insulator system through proximity. Cooper pairs from the superconductor diffuseninto the bulk of the Weyl semi-metal/topological insulator giving rise to induced superconductivity.}\label{geometry}
	\end{figure}

Similarly, one might  expect that the introduction of superconductivity in the
Weyl semi-metal would also lead to new phenomena. A heterostructure of topological insulators
and $s$-wave superconductors was studied by Meng and Balents\cite{mengbalents} who showed
that superconductivity split the Weyl modes into Boguliobov-Weyl modes. By studying vortices
in some of these phases, characterised by different number of Weyl modes, they found zero-energy
Majorana modes under some conditions. Cho {\it et al}\cite{choetal} studied superconducting states
of doped inversion symmetric Weyl semi-metals and showed that the finite momentum FFLO pairing
state  is energetically favoured over the even parity BCS state. 
Recently, Lee {\it et al}\cite{klee} studied the proximity effect in topological insulators, when the 
chemical potential is close to, but not in the bulk gap. They found that the superconducting gap penetrates 
the bulk and is observable at the naked surface, opposite to the one in proximity to the superconductor. 

However, there has been no systematic study of  proximity induced superconductivity in Weyl semi-metals,
which is the main focus of this paper. Since our model also includes the topological insulator
phase for some region of parameter space, we also provide results for proximity induced 
superconductivity for topological
insulators in
our model for comparison. Our model consists of a 3D topological insulator, converted to a Weyl semi-metal, by
including either parity breaking terms or time-reversal breaking terms or both. Time-reversal breaking 
leads to  Weyl nodes  at the same energy and surface states, which form a Fermi arc between the nodes, 
 whereas parity breaking leads to Weyl nodes 
at different energies and no Fermi arcs. 
The dispersion of the Fermi arcs or surface states is flat in a particular direction and chiral in other direction
and points along the Fermi arc can be understood as the edge states of a two dimensional Chern insulator. 
When inversion symmetry is broken however, the Weyl nodes
are separated in energy and we do not get surface states, distinct from the
bulk. Since we are mainly interested in the new physics coming from the Fermi arc states, 
unless otherwise specified, in this paper we always consider Weyl semi-metal phase induced by having a time reversal breaking perturbation.

Superconductivity is then 
induced in the semi-metal by coupling it to an $s$-wave superconductor on one of its surfaces. 
We  compute the self-energy of the topological insulator/semi-metal electrons by integrating out the superconductor
degrees of freedom and use the imaginary part of the Green's function to compute the local density of states (LDOS).
We find that the superconductor induces coherence peaks on the LDOS of the 
electrons on a few layers close to the interface and we contrast the behavior of the LDOS on
different layers for the TI and the WSM.  For the TI, we find the reduction in the LDOS close to 
$\omega=0$ which is the hallmark of the gap formation, whereas for the WSM, we find that the enhancement of the density
of states without a superconductor (the hallmark of the flat band) split into two bands with a reduction
of the density of states at $\omega=0$.  We study in detail the band structure  of the surface states 
and find that the surface state of the TI is completely gapped by the proximity effect, whereas
the surface states of the WSM get split and acquire a small gap, 
but the Weyl nodes remain unaffected.  
Thus the surface band acquires a superconducting gap in a TI, but 
no true superconducting gap is induced in WSM~\cite{mengbalents}.

We also study the behavior of the induced pairing amplitudes (singlet, triplet, 
intra-orbital and inter-orbital) as a function of the various parameters in the theory.
 As shown in Ref.~\onlinecite{balatsky}, the induced pairing amplitudes in $\text{Bi}_2\text{Se}_3$ - type
materials with tetragonal symmetry are classified in terms of the irreducible representations $\Gamma$ of the $D_{4h}$ group. 
Since we are only considering proximity with an $s$-wave conductor in this paper, we are only interested in representations with total
angular momentum $J_z=0$.
We find that the induced pairings fall of exponentially fast away from the interface both
in the TI and in the WSM. But they are not very sensitive to other parameters such as the
chemical potential and the time-reversal breaking parameters. The induced pairings increase as a function
of the superconducting pairing amplitude of the superconductor and the coupling of the superconductor to  the TI/WSM.
It is also perhaps worth mentioning that both in the singlet and triplet amplitudes, the symmetries of the
two largest amplitudes reverse between the TI and the WSM, with inter-orbital pairings being larger
in the TI and intra-orbital pairings being larger in the WSM.
\begin{figure}
  \includegraphics[width=0.45\textwidth]{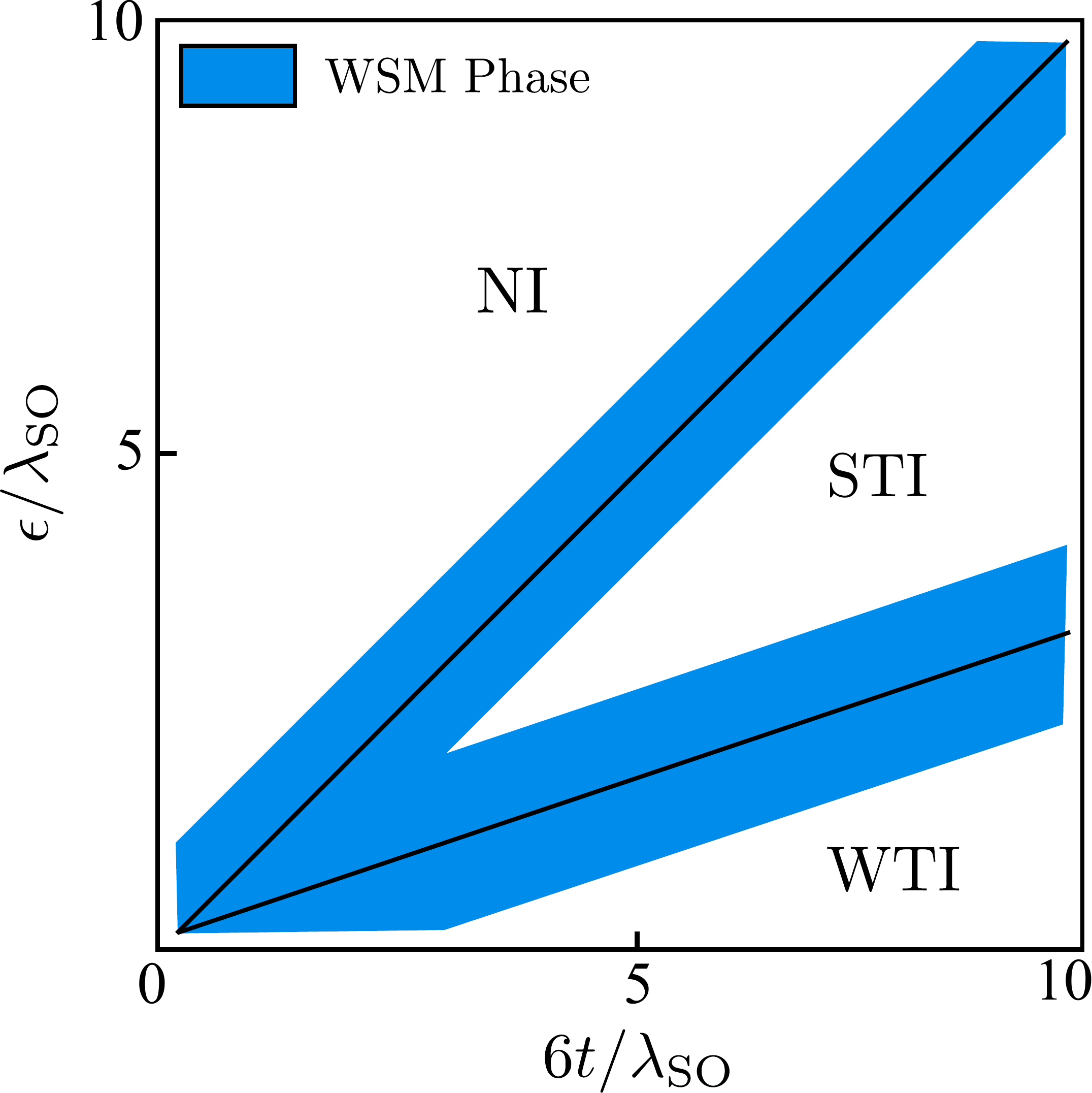}
  \caption{(Color online) A typical phase diagram of our model system. The Weyl semi-metal (WSM) phase appears at the strong topological insulator (STI)/ normal insulator (NI) ($\epsilon = 6t$) and strong topological insulator (STI)/ weak topological insulator (WTI)  ($\epsilon = 2t$) boundaries with broken time reversal/ parity perturbations. The WSM phase extends with increasing perturbations (blue/filled region). Parameters used here are $\lambda_{z} = \lambda_{\text{SO}}$, and $\bs{b} = (0.6\lambda_{\text{SO}} ,0,0)$. }
   \label{fig:phasediagram1}
\end{figure}

\section{Model System}
We start with a simple tight-binding four-band lattice model for the topological insulator (TI) in three dimensions (3DTI), which can describe strong and weak topological insulators, 
Weyl semi-metals and ordinary insulators depending on the 
parameters of the model. 
The $\text{Bi}_2\text{Se}_3$ family of 3DTI, have an effective
description in terms of the Hamiltonian given by 
$H_0 =  H_{\text{C}}+ H_{\text{SO}}$ \cite{vazifehfranz} with 
\begin{align}\label{eq:h0}
 H_{\text{C}}&=\epsilon \sum_{\bs r} \psi^{\dagger}_{\bs r}\tau_x\psi_{\bs r} - t\sum_{\langle \bs r, \bs r' \rangle}\psi^{\dagger}_{\bs r}\tau_x\psi_{\bs r'} + ~\text{h.c.} \nonumber \\
  {\rm and} ~~~ H_{\text{SO}} &= i\lambda_{\text{SO}}\sum_{\bs r} \psi^{\dagger}_{\bs r}\tau_z\left( \sigma_x\psi_{\bs r + \bs y} - \s_y\psi_{\bs r + \bs x} \right) \nonumber \\
 &+i\lambda_z\sum_{\bs r}\psi^{\dagger}_{\bs r}\tau_y\psi_{\bs r + \bs z} +  ~\text{h.c.}
\end{align}
where $\psi_{\bs r}$ is the fermion operator in TI region. $\bs r, \bs r'$ refer to site indices in all three dimensions (in the TI region) 
and $\bs r + \bs x$ refers to the nearest neighbour of site at $\bs r$ in $x$ direction (similarly for $y$ and $z$ directions). 
Here $z$ is taken as the growth direction and $\bs{\s}$ and $\bs{\tau}$ denote Pauli matrices in spin and parity (orbital) space respectively.
$\epsilon$ and $t$ denote the on-site and nearest neighbour hopping amplitudes.
$\lambda_{\mathrm{SO}}$ and $\lambda_{z}$ are the (possibly anisotropic) spin-orbit (SO) interaction strengths 
in the $x$-$y$ plane and in the $z$ direction respectively. 

 The  topological invariants for the  3DTI, $\nu_{\mu} = (\nu_0; \nu_1, \nu_2, \nu_3)$ can be computed easily  
(due to parity invariance \cite{fukane}) and are given by
\begin{align}
 (-1)^{\nu_0} &= \text{sgn}\left[ (\epsilon - 6t) (\epsilon + 6t) (\epsilon - 2t)^3 (\epsilon + 2t)^3 \right], \non \\
(-1)^{\nu_i} &= \text{sgn}\left[ (\epsilon + 6t) (\epsilon - 2t) (\epsilon + 2t)^2 \right], \nonumber
\end{align}
for $i = 1, 2, 3$. This implies that we have the following phases:
\begin{align}  \begin{array}{ccc}
 |\epsilon| > |6t| & \nu_{\mu} = (0;0,0,0) & \text{Ordinary Insulator} \\
 |6t| > |\epsilon| > |2t| & \nu_{0} =1 & \text{Strong TI} \\
 |2t| > |\epsilon| > 0 & \nu_{0} = (0;1,1,1) & \text{Weak TI}\end{array} 
\end{align}

At the boundaries of the topological phase transitions (at $\epsilon \approx \pm 6t, \pm2t$), 
the bulk gap closes and the effective hamiltonian is a massless Dirac hamiltonian. 
By introducing either parity (inversion) or time reversal (TR) symmetry breaking perturbations  
to the Hamiltonian $H_0$, the Dirac node can be split in 2 weyl nodes separated in energy or momentum respectively. 
Thus the Hamiltonian for the WSM is given by 
 $H_W = H_0 + H_{\text {E}}$  where 
\begin{equation}\label{eq:he}
 H_{\text{E}} = \sum_{\bs r} \psi^{\dagger}_{\bs r} \left( b_0\tau_y\s_z - b_x\tau_x\s_x + b_y\tau_x\s_y + b_z\s_z\right)\psi_{\bs r} ~.
\end{equation}
Here $b_0$ and ${\bs b}$  are parameters that break inversion and TR symmetry
respectively \cite{vazifehfranz}. 
Note that the Dirac node mentioned here is the 3D Dirac node that occurs in the bulk spectrum at the phase transition between the normal and
topological insulator and should not be confused with the 2D Dirac nodes, which occur in the surface spectrum of the TI phase. The phase diagram of the different phases in this model is given in Fig.~\ref{fig:phasediagram1}. 

\begin{figure}[ht]
\includegraphics[width=0.48\textwidth]{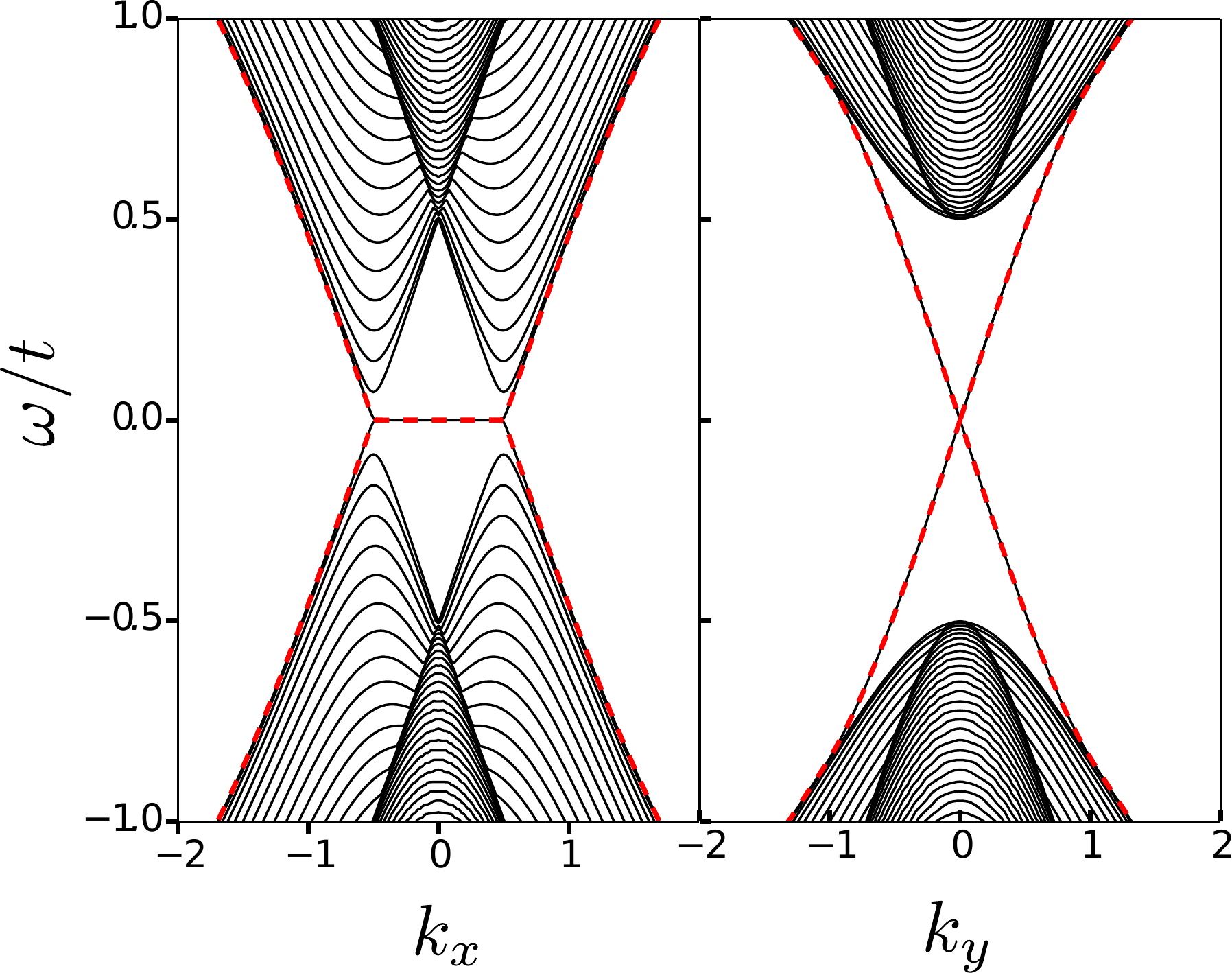}
\caption{(Color online) The  dispersion for a WSM with 2 Weyl nodes at $\pm b_x /\lambda_z$ is shown  along  $k_x$ and  $k_y$.
The parameters used are $\bs b =(0.50t,0,0), \lambda_{SO}=\lambda_z=0.50t$. The dashed (red) lines denote both the surface bands. Note that the surface states at opposite ends have opposite chirality.}
\label{dispersion}
\end{figure}

\begin{figure*}[tb]
\includegraphics[width=1.\textwidth]{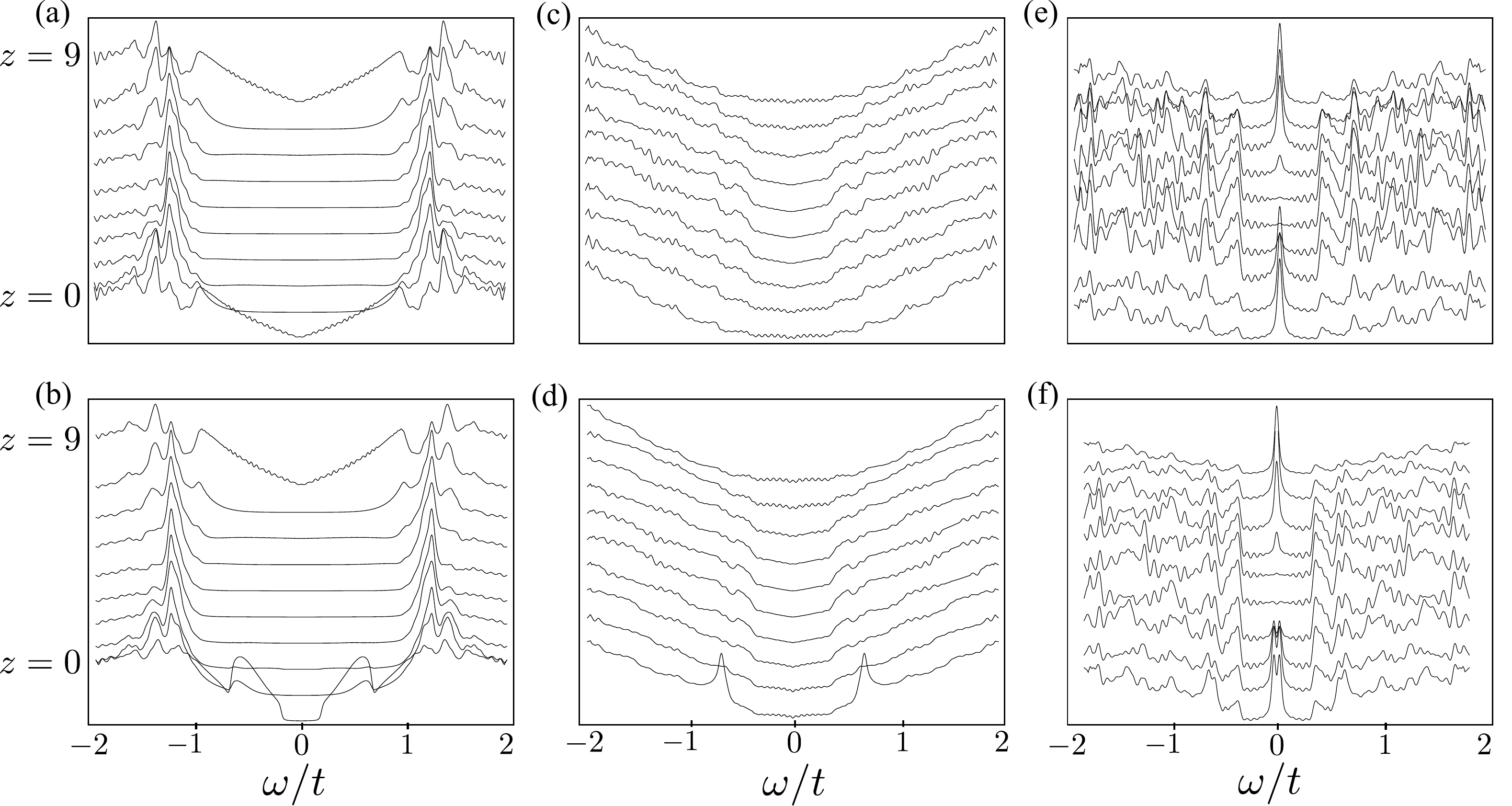}
\caption{This panel shows the LDOS 
for  the TI and the WSM with the top row without coupling to the superconductor and the bottom row with coupling to the superconductor.
(a) and (b) show the LDOS
for the TI integrated over all momenta  without  and with coupling to the superconductor respectively.
 (c) and (d) show the same for the WSM. (e) and (f) show the LDOS at $k_y=0$ summed over all $k_x$
 for the WSM  without and with coupling to the superconductor respectively.
 The LDOS for the different layers are vertically offset for visibility.
The parameters used are $\lambda_{SO} = \lambda_z = 0.5t$, $\Delta = 0.7t$, $\epsilon = 4t$ (for TI) or $\epsilon = 6t$ (for WSM) and $\bs b = (0.5t,0,0)$ for WSM.}
\label{coherencepeak}
\end{figure*}

For finite systems, both TI and WSM phases give rise to  surface states. For strong topological insulators, 
surface states exist on  each surface as mid-gap states\cite{TI}, whereas in weak topological insulators, 
surface states arise only on  particular surfaces depending on the values of $\nu_i$ ($i=1,2,3$)\cite{weakTI}. 
For  Weyl semi-metals (within this model), if only inversion symmetry is broken, 
the Weyl nodes are separated in energy and there are no surface states 
separable from the bulk states. Surface states arise only when TR symmetry is broken. As an example, if the TR symmetry is broken by ${\bs{b}} = b_x \hat{x}$, then
the Weyl nodes occur with a separation of $b_x/\lambda_z \hat{x}$ in momentum space. 
In this case, we find surface states exist on the surfaces parallel to the $x-y$ and $x-z$ plane (not on the third plane).  
For a large enough system, the dispersion of surface states is flat between the two Weyl nodes along the $k_x$ direction,  and is linear
along $k_y$ (on $x-y$ plane) or $k_z$ (on $x-z$ plane). 
The surface states exist only between the two Weyl nodes and 
the states on opposite surfaces have opposite chiralities, as illustrated in Fig.~\ref{dispersion}. These are the Fermi arc states.
Fermi arcs can be understood as the edge states of a Chern insulator that exists for each value of $k_x$ between the Weyl nodes. 
It  may also be worth noting that the WSM formed near $\epsilon \sim 6t$ and $\epsilon \sim 2t$ have their  chiralities
reversed for the top and bottom edges.

For the rest of this paper, we concentrate on the TR broken Weyl semi-metal, with $b_0=0$ and ${\bs{b}} \ne 0$, 
since we are interested in the proximity
effect of the superconductor on the surface, i.e.,  on the Fermi arc states.

\section{Coupling to the superconductor}

We now couple one of the surfaces of the WSM to an $s$-wave superconductor as shown in Fig.~\ref{geometry}. The bulk Hamiltonian of the $s$-wave superconductor is given by 

\begin{align}\label{eq:hs}
 H_\text{S} &= \epsilon_\text{sc} \sum_{\bs R,\sigma} \Phi_{\bs R,\sigma}^{\dagger}\Phi_{\bs R,\sigma} - 
 t_\text{sc} \sum_{\langle \bs R, \bs R' \rangle,\sigma} \Phi_{\bs R,\sigma}^{\dagger}\Phi_{\bs R',\sigma} \nonumber \\
		&+ \sum_{\bs R} \Delta \Phi_{\bs R,\uparrow}^{\dagger}\Phi_{\bs R,\downarrow}^{\dagger} + ~\text{h.c.}~,
\end{align}
where $\Phi$ is the fermion operator in the superconductor. $\bs R, \bs R'$ refer to site index in all three directions (in the superconducting  region).
$\epsilon_\text{sc}$ and $t_\text{sc}$ denote the on-site energy and nearest-neighbour hopping amplitudes in the superconductor respectively. 
The coupling is a tunnelling term between the top layer of the superconductor and the bottom layer of the WSM - 
\begin{align}\label{eq:ht}
 H_{\text{T}} = \sum_{\bs r_c, \tau, \sigma} \tilde{t}_{\tau} \psi_{\bs r_c,\tau,\sigma}^{\dagger}\Phi_{\bs r_c+\bs z,\sigma}+ ~\text{h.c.}~,
\end{align}
where, $\bs r_c$ denotes the sites in the last layer of the WSM (perpendicular to the interface) and $\bs r_c + \bs z$ denotes the first layer of 
the superconductor. 
$\tau$ denotes the orbital in the WSM and $\sigma$ is the spin. 
$\tilde{t}_\tau$ is the tunnelling amplitude which can be different for the two orbitals.
In this work, we have assumed that $t_\tau$ is same for both orbitals for simplicity.
Using different tunnelling  amplitudes for the two orbitals will change the results quantitatively, but not qualitatively. 
For detailed results as a function of the ratio of the tunnelling amplitudes for the TI, see \onlinecite{balatsky}.

\begin{figure}[h]
\includegraphics[width=0.48\textwidth]{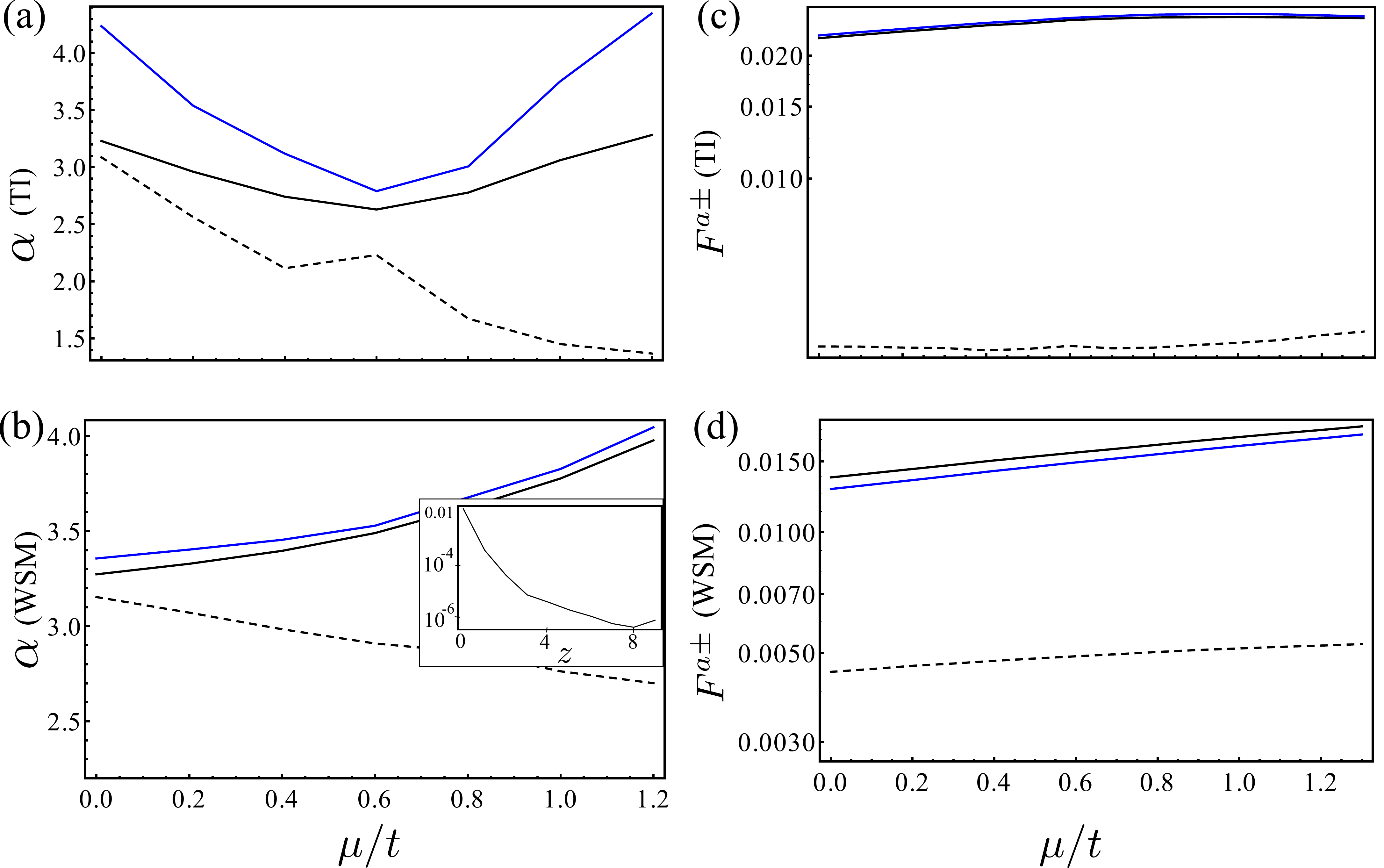}
\caption{(Color online) (a) and (b) show the exponential parameter $\alpha$ of the decay (in the unit of 1/lattice spacing in $z$) for various cases in the TI/WSM,
as a function of
the chemical potential $\mu/t$. They are  
denoted by the black and blue (gray) solid lines for $s$-wave intra and inter--orbital amplitudes respectively,  and the black line for $p$-wave intra--orbital amplitude.  In the inset of (b), we show a typical exponential decay of the pairing in the bulk. The case shown here is  the decay of intra-orbital $s$-wave pairing in 
the WSM.  (c) and (d) show the almost flat behavior of the pairing amplitudes $F^{a \pm}$ as defined in Eq.~(\ref{eq:F}) at $z=0$  as a function of the
chemical potential.  
The various parameters used are $\lambda_{\text{SO}}=\lambda_z=0.5 t$, $\Delta=0.7t$, $\lambda_S = 0.3t$, $\epsilon = 6t, ~ \bs{b}=(0.5t,0,0)$ for WSM and $\epsilon = 4t, ~ \bs{b}=(0,0,0)$ for TI.}
\label{fig:withz}
\end{figure}

\begin{figure}[h!t]
\includegraphics[width=0.4\textwidth]{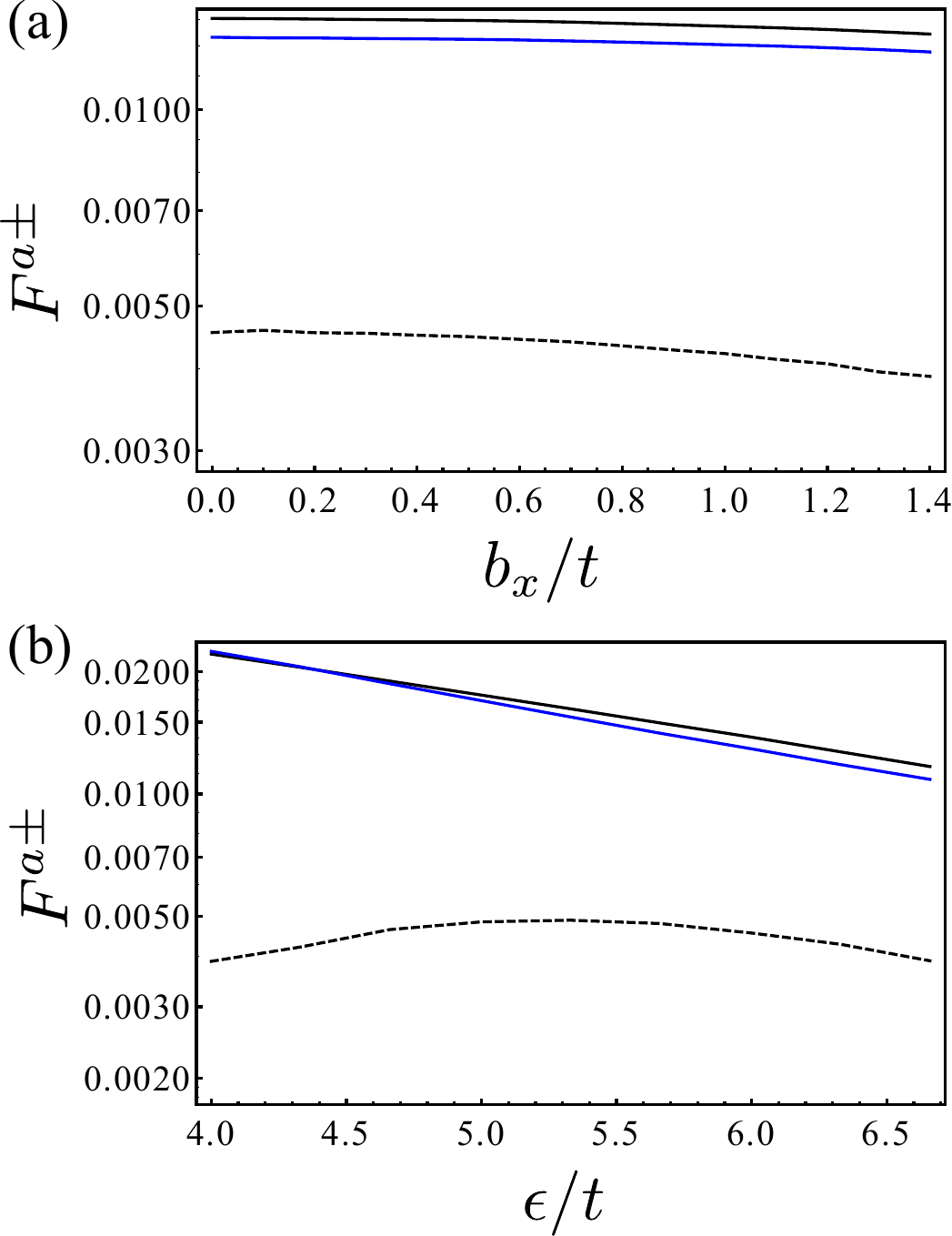}
\caption{(Color online) The pairing amplitudes as a function of  (a) the time-reversal
breaking parameter $b_x/t$  (b) $\epsilon/t$.  The different pairing amplitudes are denoted by the black and blue (gray) solid lines as $s$-wave intra--orbital and inter--orbital amplitudes respectively, and the black line as $p$-wave intra--orbital amplitude. Note the crossing of the black and  blue lines (crossing of intra and inter orbital pairings) as a function of $\epsilon/t$.
Various parameters used are $\lambda_{\text{SO}}=\lambda_z=0.5 t$, $\Delta=0.7t$, $\lambda_S = 0.3t$. For (a), we have used $\epsilon = 6t$, and $\bs{b}=(0,0,0)$ for (b).}
\label{WSMparameter}
\end{figure}

The Nambu basis for the fermions in the WSM, denoted by $\p^{\dagger}_{\bs r}$ is given by
\begin{align}
  \left( 
     \psi^{\dagger}_{\bs r, \uparrow, 1}, \psi^{\dagger}_{\bs r, \downarrow, 1},\psi^{\dagger}_{\bs r, \uparrow, 2}, \psi^{\dagger}_{\bs r, \downarrow, 2}, \psi_{\bs r, \downarrow, 1},
       -\psi_{\bs r, \uparrow, 1},   
     \psi_{\bs r, \downarrow, 2}, -\psi_{\bs r, \uparrow, 2} \right) \label{eq:nambu}
\end{align}
where  $\uparrow,\downarrow$ refer to the spin and $1,2$ refer to the orbitals. 
Since the Hamiltonian is quadratic in the superconductor degrees of freedom $\Phi$, we can integrate
them out and compute an effective action for the WSM.
Following the analysis for the self-energy in Ref.\cite{simonbena}, we decouple the superconductor and TI degrees of freedom and define the 
(Nambu-Gorkov) Green's function $G(\omega)$ for the WSM. The derivation is sketched in Appendix A. The Green's function is - 
\begin{align} G(\omega)  = \left[(\omega + i\delta)\mathbf{I} - H_W - \Sigma(\omega)\right]^{-1}.
\end{align}
where the self-energy is - 
\begin{align}\label{eq:GF}
\Sigma_{\bs r} (\omega) = \delta_{\bs r, \bs r_c} \frac{\pi N(0) (\tilde{t})^2}{\sqrt{\Delta^2 - \omega^2}}	
\left[ \omega \mathbf{I}_{\eta} - \Delta \eta^x \right] \left(\mathbf{I}_{\tau} + \tau^x \right)
\mathbf{I}_{\sigma} 
\end{align}
with $\bs{r}_c$ denoting the sites in the last layer of WSM. 
Note that the $G$ at each site is an $8\times 8$ matrix comprising of the spin, orbital and particle-hole pseudo-spin
subspaces. Here, we have used only the local (on-site) component of the self energy. 
This  approximation usually works very well and we shall justify this in the last section by comparing our
results with this approximation to the result obtained using  exact diagonalisation.


\begin{figure*}[tb]
\includegraphics[width=1.\textwidth]{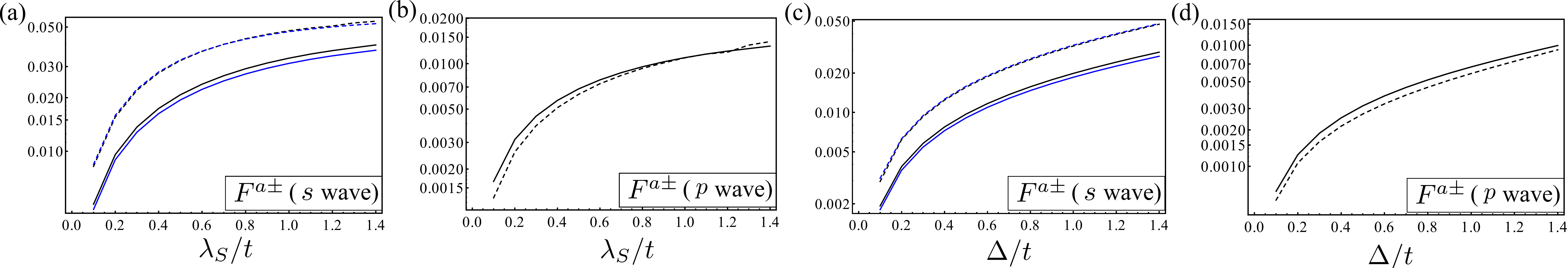}
\caption{(Color online) We depict the pairing amplitudes $F^{a \pm}$ as defined in Eq.~(\ref{eq:F}) at $z=0$ for the intra- and inter-orbital $s$-wave and intra-orbital $p$-wave  
channels as a function of the various parameters  in the model. The black and blue (gray) solid lines denote  intra- and inter-orbital pairing for WSM respectively, whereas, the black and blue (gray) dashed lines mark intra and inter-orbital pairing for TI respectively. (a) and (b) show the increase in both the $s$-wave and $p$-wave pairing channels as a function of the coupling $\lambda_S$ between the superconductor and the WSM/TI. (c) and (d) show the increase as a function of the superconducting $s$-wave pairing amplitude $\Delta$ in the superconductor.  Various parameters used, which are not varied, are $\lambda_{\text{SO}}=\lambda_z=0.5 t$, $\Delta=0.7t$, $\lambda_S = 0.3t$, $\epsilon = 6t, ~ \bs{b}=(0.5t,0,0)$ for WSM and $\epsilon = 4t, ~ \bs{b}=(0,0,0)$ for TI.}
\label{pairingmag}
\end{figure*}

\section{LDOS and the pairing amplitude}

In this section, we use the Green's function to obtain the local density of states (LDOS) and the induced  pairing amplitude  both in the TI and WSM phases and discuss the dependence of the pairing on the various parameters of the model.

Using the Green's function that we derived in the previous section, we can compute the local density of states (LDOS) in the TI/WSM
using 
\begin{align}
D(\omega,\bs{r}) &= -\frac{1}{\pi} \sum_{\sigma,\tau}{\rm Im} G^{\sigma\tau}_{\bs{r}\bs{r}}~,
\end{align}
$\sigma$ and $\tau$ being the spin and orbital index. As we increase the coupling to the superconductivity, the LDOS shows the appearance of coherence peaks in the band structure as a signature of proximity induced superconductivity. This is shown  in the panel in Fig.~\ref{coherencepeak}  
where we have plotted the LDOS at different values of $z$,  the 
layer index,  both for the TI and the WSM.  Panels (a) and (b) show the 
LDOS as a function of energy (summed over all momenta) for the TI.  Note the Dirac spectrum feature of the surface states for the TI
(layers $z=0$ and $z=9$). We can also  clearly see the dip in the density of states and the appearance of the coherence peaks in the first 2 layers.  
The coherence peaks are not sharp because we are at zero doping. (For the TI, the sharpness of the coherence peaks increases with
the doping, due to the increase in density of states.  For WSM, we have checked that there is no significant change when the doping
is increased, because there is a significant density of states at the Weyl node even at zero doping). Panels (c) through (f) are for the Weyl semi-metal, with (c) and (d) being the LDOS summed over all momenta and (e) and (f) being the LDOS at $k_y=0$ summed over all $k_x$.  First, we note  the absence of the Dirac spectrum feature in the edge states. Instead, there is peak in the DOS at $\omega =0$ (and $k_y=0$)  which is the signature of the flat band  in the absence of coupling to the superconductor. (This feature gets lost when all $k_y$ is summed over, which is
why we have also chosen to show the band-structure without summing over $k_y$.)  With the coupling to the superconductor,
the single peak  splits into two with a small gap. The effect of the proximity of the superconductor on the surface states will be studied in greater detail in the next section.

Following \cite{sigrist}, we now define the different induced pairing amplitudes. 
Assuming translation invariance in $x$ and $y$ directions, 
we go to momentum space $\bs k = (k_x, k_y)$ in 2 directions. For each momentum ${\bs k}$ and each $z$ coordinate, 
the Green's function can be written as
\begin{align}
G_{z,\bs{k}}(\omega) &=\left(\begin{array}{cc}
  h_{z}(\omega, \bs{k}) & \bar{\Delta}_z(\omega,\bs{k}) \\
  \bar{\Delta}^{*}_z(\omega,\bs{k}) & h'_{z}(\omega, {\bs k})
 \end{array}\right)~.
\end{align}
The $4\times 4$ pairing matrix $\bar{\Delta}_{z;\sigma\tau,\sigma'\tau'}(\omega,{\bs{k}}) $ is  related to the
pairing amplitudes as 
\begin{align}
\bar{\Delta}_{z;\sigma\tau,\sigma'\tau'}(\omega,{\bs{k}}) =& \int_0^{\infty} \frac{dt}{2\pi}\langle c_{z,{-\bs{k}},\sigma,\tau}(t) 
c_{z,{\bs{k}},\sigma',\tau'}(0)\rangle e^{i\omega t} \nonumber  \\
=& \int_0^{\infty} \frac{dt}{2\pi}{\hat\Delta}_{z;\sigma\sigma'\tau\tau'}({\bs k}, t)e^{i\omega t},
\end{align}
where the last equality defines ${\hat\Delta}_{z;\sigma\sigma'\tau\tau'}({\bs k},t)$. 
We only consider the equal time ($t=0$) pairing amplitudes from here on. 
It is useful to form combinations of the pairing amplitudes 
which are even $(+)$ or odd $(-)$ under exchange of orbital index as - \\
\begin{align}
\hat{\Delta}^{i \pm}_{z;\sigma\sigma'} ({\bs{k}}) =& \hat{\Delta}_{z;\sigma\sigma', 11}({\bs k}) \pm \hat{\Delta}_{z;\sigma\sigma', 22}({\bs k}) \nonumber \\
 \hat{\Delta}^{I \pm}_{z;\sigma' \sigma} ({\bs{k}}) =& \hat{\Delta}_{z;\sigma\sigma', 12}({\bs k}) \pm \hat{\Delta}_{z;\sigma\sigma', 21}({\bs k}) 
\end{align}
 where the superscript $i$ or $I$ on the LHS refers to intra-orbital and inter-orbital pairings. 
Each of these $\hat{\Delta}$ is a $2 \times 2$ matrix in spin space and can be written as a sum of singlet and triplet components - 
\begin{align}
\hat{\Delta} (\bs k) = i \sigma^y \psi({\bs k}) + i \sigma^y (\mathbf{d}({\bs k}) \cdot \boldsymbol{\sigma} ) 
\end{align}
where due to Fermi statistics, $\psi({\bs k})$ is even and $\mathbf{d}({\bs k})$ is odd under $ {\bs k} \rightarrow -{\bs k}$ for $i\pm$ and $I+$ pairings, 
whereas, $\psi({\bs k})$ is odd and $\mathbf{d}({\bs k})$ is even under $ {\bs k} \rightarrow -{\bs k}$ for $I-$ pairings. Then, the even and odd intra-orbital and even interorbital amplitudes have the usual s-wave spin singlet and p-wave spin-triplet pairing 
while the odd interorbital amplitude has a p-wave spin singlet and s-wave spin triplet pairing. Reference \cite{balatsky} found that an s-wave superconductor does not induce even frequency odd interorbital pairings in TI (although a p-wave superconductor can do so), however, 
it does induce odd frequency odd interorbital pairings.  In this work, since we are only interested in equal time correlations,
we ignore the even frequency odd interorbital and all odd frequency pairing amplitudes.

The Hamiltonian of  $\text{Bi}_2\text{Se}_3$ - type
material considered here has tetragonal symmetry (since it is written on a cubic lattice) and the induced pairings 
are classified in terms of the irreducible representations $\Gamma$ of the $D_{4h}$ group \cite{balatsky}.
Thus, $\psi({\bs k})$ and $\mathbf{d}({\bs k})$ must have a 
functional dependence on ${\bs k}$, which forms an irreducible representation of $D_{4h}$. 
Since we are only considering proximity with an $s$-wave superconductor in this paper and assuming that the angular momentum in the
${\hat z}$ direction is conserved in the 
tunneling process, there are only three  relevant representations - $A_{1g}$, $A_{1u}$ and $A_{2u}$. 
Upto linear order in ${\bs k}$ (and taking $k_z = 0$) we have, 
$\psi({\bs k}) = 1$ for $A_{1g}$, $ \mathbf{d}({\bs k}) = (k_x, k_y, 0) $ for $A_{1u}$ and $ \mathbf{d}({\bs k}) = (k_y, -k_x, 0) $ for $A_{2u}$. 
On a square lattice, we may replace ${\bs k}^2$ by $1 - \cos ({\bs k})$ and terms linear in ${\bs k}$ by $\sin ({\bs k})$. 
We can then classify the   $ \hat{\Delta} ({\bs k}) $ found numerically, 
by finding their inner product with the basis functions. For this, we define 
\begin{align}
F^{a \pm}_{\sigma \sigma'} &= \frac{1}{2 N_{\bs{k}}} \sum_{{\bs k}} S^{*}_{\sigma \sigma'} ({\bs{k}}) \hat{\Delta}^{a \pm}_{\sigma' \sigma} ({\bs {k}})
\end{align}
where, $S_{\sigma \sigma'} ({\bs k})$ is one of the basis functions given above. The superscript $a=i,I$ refers to intra and inter-orbital pairing 
respectively. Then the inner product is 
\begin{align}\label{eq:F}
F^{a \pm} = \sum_{\sigma \sigma'} F^{a \pm}_{\sigma \sigma'}~. 
\end{align}
We find that the spin singlet amplitudes have a dominant component with $A_{1g}$ pairing as expected, but the triplet components have $A_{2u}$ pairing and not $A_{1u}$  in both TI and WSM\cite{note1}.  This is due to the form of the spin-momentum locked low energy Dirac surface state which enforces
the vanishing of the $A_{1u}$ triplet amplitude\cite{balatsky}.  

We note that for the spin singlet, the odd intra-orbital pairing is lower by two orders of magnitude compared to the even orbital pairings. 
Thus, only s-wave spin singlets with even orbital pairings are dominant. 
For spin triplet, the even orbital pairings are lower by two orders of magnitude with respect to the odd intra-orbital pairing. 
Thus, a p-wave spin triplet with odd intra-orbital pairing is dominantly induced.
Hence, for both the TI and the WSM, we only display the behavior of the following three amplitudes - spin singlet even intra- and inter-orbital 
pairing and spin triplet odd intra-orbital pairing. 

In both the TI and the WSM, the pairing amplitudes fall off exponentially in the bulk\cite{stanescu,klee}. The falloff can be numerically fit
to  an exponential $F\propto e^{-\alpha z}$, where the direction $z$ is perpendicular to the surface of contact to the superconductor. In Fig.~\ref{fig:withz}(a),(b)we show how $\alpha$ varies for the various pairing amplitudes as a function of the chemical potential $\mu$ for TI and the WSM. We note that $\alpha$ starts to decrease as we increase $\mu$ in the case of TI, both for $s$ and $p$ wave amplitudes, which means a greater penetration~\cite{klee}. 
 For the case of WSM, the $s$-wave pairing has a decreased penetration with increasing $\mu$ in contrast to the TI, whereas the $p$-wave pairing
 has mildly increasing penetration.

In Fig.~\ref{fig:withz}(c),(d) we compare how the various pairing amplitudes at $z=0$ for the TI and the WSM change as a function of 
the chemical potential $\mu/t$. We note that the spin singlet  amplitudes are  higher than the spin triplet amplitudes in all cases. There is also
not much variation between the TI and the WSM as far as the the spin singlet amplitudes are concerned.
But the spin triplet amplitudes have  substantially larger   variation  between the TI and the WSM.

We have also studied the behavior of these pairing amplitudes  in both the TI and the WSM  as a function of various other parameters. 
The pairing amplitudes remain flat with time-reversal symmetry breaking perturbations $b_x$, as shown in Fig.~\ref{WSMparameter}(a). 
In Fig.~\ref{WSMparameter}(b) $\epsilon/t$ parameterizes the flow from the TI to the WSM, which shows a switching from inter to intra--orbital pairing as one moves from 
the TI to the WSM. We show the pairing amplitudes in Fig.~\ref{pairingmag}  at $z=0$ as a function of  the  coupling
 to the superconductor $\lambda_S = \pi N(0)({\tilde t})^2$ and 
as a function of the the superconducting pairing amplitude  $\Delta$ and in general, we see that all the pairing amplitudes increase  as the parameters
increase.

\section{Surface states \& comparison with exact diagonalization}

\begin{figure}
  \includegraphics[width=0.45\textwidth]{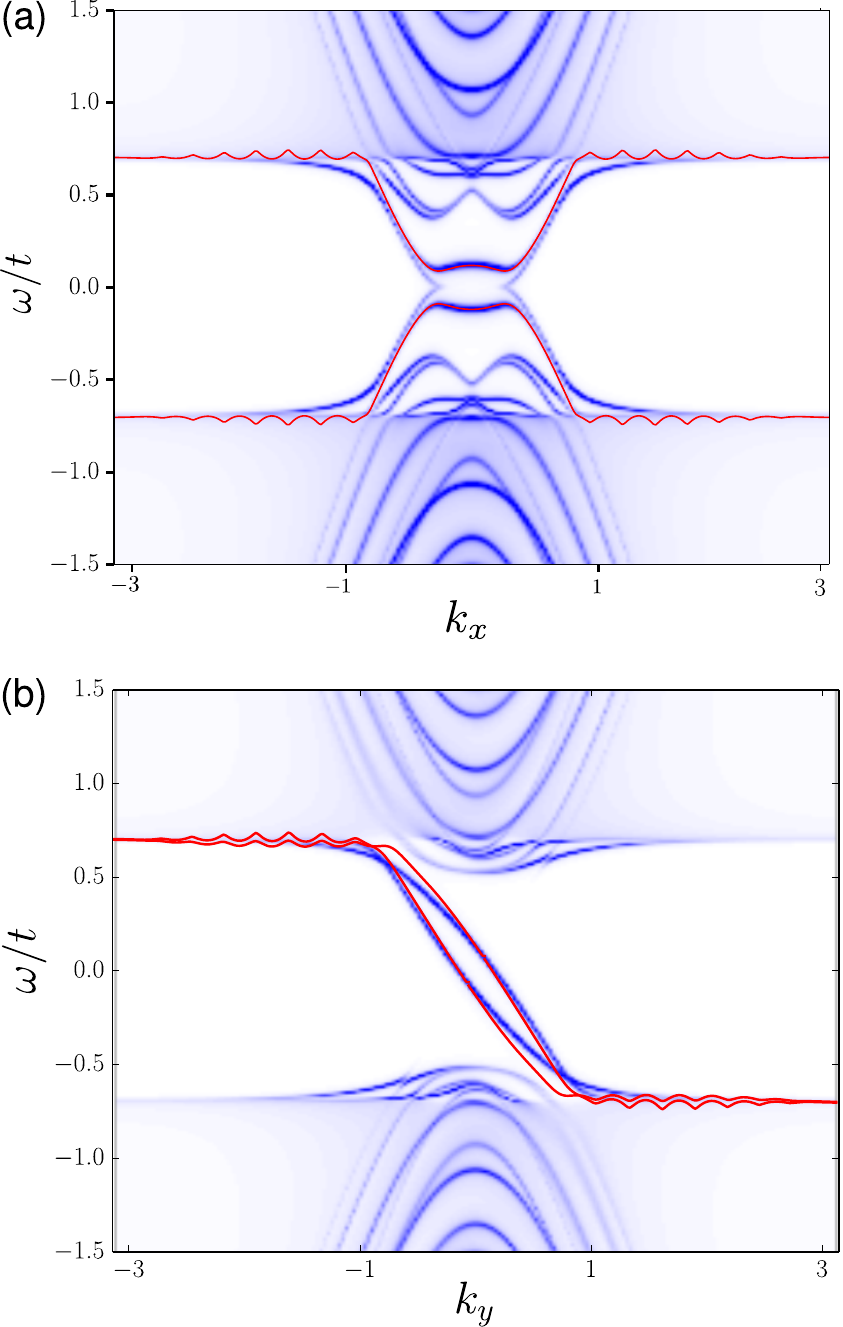}
  \caption{(Color online) Comparison of the Green's function technique that we used with exact diagonalization results for the effect of the proximity induced superconductivity in the surface bands of the WSM with momenta (a) $k_x$ and (b) $k_y$ where the Weyl nodes lie along $k_x$. The blue (gray) high density lines are the modified bands in the system with proximity to the superconductor obtained from the LDOS at $z=0$,  while the red (darker) solid line is the surface band at $z=0$ via exact diagonalization. Note that we have only depicted the surface band  at $z=0$ and have suppressed the
 other surface state. The induced gap vanishes at the Weyl nodes for a large enough system size, but the surface band splits. The
 various parameters that have been used for the LDOS are $\lambda_{\text{SO}}=\lambda_z=0.5 t$, $\Delta=0.7t$, $\lambda_S = 0.9t$, $\epsilon = 6t, ~ \bs{b}=(0.5t,0,0)$ and the number of sites in $z$ has been taken to be  20. We have used  $k_y = 0$ for (a) and  $k_x = 0$ for (b).}
   \label{fig:surface}
\end{figure}
Finally, we discuss the effect of superconducting proximity on  the surface states and we compare the results from the Green's function method with an exact diagonalization. After computing the self-energy of the electrons in the WSM due to the proximity effect  as in Eq.~(\ref{eq:GF}), we  can construct the effective band-structure by solving 
\begin{align}\label{eq:eigval}
\mathrm{Det}\left[ H_{\mathrm{W}} + \Sigma(\omega) - \omega \right] = 0~,
\end{align}
which, in turn, is the equivalent of finding  the peaks in the LDOS of the system. For the case of the TI, the surface state acquires an induced pairing which gaps the surface band completely in agreement with earlier results\cite{stanescu}, whereas  for the  surface state of the WSM,  the induced gap is
much smaller and actually vanishes at the Weyl nodes.  This is shown in Fig.~\ref{fig:surface}(a) and (b), where the effective 
band structure of the WSM 
in proximity with a superconductor  is plotted
as a function of $k_x$ and $k_y$ respectively. The same band structure has also been plotted using exact
diagonalisation. Note that Fig.~\ref{fig:surface} is only the bandstructure at the surface in proximity to the superconductor. This is why the other surface state, 
with opposite chirality is not visible here. 
The first point to note is that there is no qualitative difference in the band structure using the Green's function technique
and using exact diagonalisation. This clearly justifies the approximation of  using only the on-site or local component in
the self-energy.
Also, note that in contrast to the dispersion in Fig.~\ref{dispersion} without the proximity effect, we see here that the proximity
to the superconductor has split the flat band into two,  giving rise to a small anisotropic gap. However, the states at the Weyl nodes are not gapped.
This  is not unexpected because the $s$-wave
superconducting correlations couples the electrons at one node of a certain chirality to holes at the other node of the {\it same} chirality (because
the two nodes have opposite chirality, but the holes and electrons also have opposite chirality), hence, no gap can open up\cite{mengbalents}.
It is also of interest to consider the band structure as a function
of $k_y$ as shown in Fig.~\ref{fig:surface}(b). The edge states of the Chern insulator for each fixed value of  $k_x$ between the nodes,  
are now  split by the proximity effect into two edge states, each carrying {\it half} the Chern number of the original edge state.
 We also  compare the LDOS computation with an exact  diagonalization and as can be seen in the Figure, the results match quite well.

\section{Discussion and conclusions}

In summary, we have provided a detailed study of proximity induced superconductivity in Weyl
semimetals. We have focused on proximity of $s$-wave superconductor in the current work,
though a similar analysis can also be made for $p$-wave and $d$-wave superconductors.
We find that despite the presence of bulk metallic states in the WSM, the induced pairing
remains confined to a few layers close to the interface and in fact, falls off exponentially
fast away from the interface. We note that the $s$-wave superconductor induces both $s$-wave
and $p$-wave pairing, but the  induced $p$-wave pairing  is always smaller than the dominant
$s$-wave pairing.  We also  find that increasing the  chemical potential does not increase either 
the penetration into the bulk, or the ratio between the $p$-wave and the $s$ wave amplitudes
significantly. Both $s$-wave and $p$-wave components of the induced pairing can, however,
be increased by increasing the pairing amplitude in the superconductor or by increasing the
coupling to the superconductor.

\section*{Acknowledgments} U.K, S.P and S.R  would like to thank J. D. Sau  for useful discussions. Computational work for this study was carried out at the cluster computing facility in the Harish-Chandra Research Institute (http://www.hri.res.in/cluster). A.K was supported by the College of Arts and Science at Indiana University, Bloomington. Further funding was provided by the Offices of the Vice President for Research and the Vice Provost for Research at Indiana University through the Faculty Research Support Program. We would also like to thank the anonymous referee for comments, which were helpful in improving the clarity and presentation of the work.

\vspace{2cm}

\appendix

\section{Computation of the self-energy}

We compute the self-energy of the Weyl semimetal electrons tunnel-coupled to a superconductor along
one of its surfaces, following Ref.[\onlinecite{simonbena}]. The complete Hamiltonian is given as
\begin{align}
H = H_C+H_{SO}+H_E+H_{S}+H_T  \label{eq:Ah}
\end{align}
where $H_C+H_{SO}$ is given in Eq.(\ref{eq:h0}), $H_E$ is given in Eq.(\ref{eq:he}), $H_S$
is given in Eq.~(\ref{eq:hs}) and $H_T$ is given in Eq.~(\ref{eq:ht}).
We work in the Nambu basis for the fermions given in Eq.~(\ref{eq:nambu}). 
In this basis the hamiltonian for the WSM is  $H_W = H_C + H_{SO} + H_{E}$  with 
\begin{align}
& H_C =- t \sum_{\langle \bs r, \bs r' \rangle} \p^{\dagger}_{\bs r} h_{c} \p_{\bs r'} + h.c. + \epsilon \sum_{\bs r} \p^{\dagger}_{\bs r} h_{c} \p_{\bs r} 
\non\\
& H_{SO} = \sum_{\bs r} \p^{\dagger}_{\bs r} h_{sx} \p_{\bs r + \bs x} + \sum_{\bs r} \p^{\dagger}_j h_{sy} \p_{\bs r + \bs y} \non \\
  & \quad \quad \quad + \sum_{\bs r} \p^{\dagger}_{\bs r} h_{sz} \p_{\bs r + \bs z} + h.c. 
 \non \\
& H_E = \sum_{\bs r} \p^{\dagger}_{\bs r} h_E \p_{\bs r} ~.
\end{align}

\noindent
The various $h$ matrices defined above are given as
\begin{align}
& h_c = \eta^z\tau^z \mathbf{I}_{\sigma}, \non\\
& h_{sx} = \lambda_{SO} \eta^z \tau^z \sigma^y \non\\
& h_{sy} = \lambda_{SO} \eta^z \tau^z \sigma^x \non \\
& h_{sz} = i\lambda_z \eta^z \tau^y \mathbf{I}_{\sigma} \non\\
& h_E = (b_o \eta^z \tau^y \sigma^z) \non \\
& \quad +  (-b_x \mathbf{I}_{\eta} \tau^x \sigma^x  + 	b_y \mathbf{I}_{\eta} \tau^x \sigma^y + b_z  \mathbf{I}_{\eta} \mathbf{I}_{\tau} \sigma^z)~.
\end{align}
where, as mentioned below Eq.~(\ref{eq:h0}), $\bs{\s}$ and $\bs{\tau}$ denote Pauli matrices in spin and parity (orbital) space 
and $\bs{\eta}$ represents the particle-hole pseudo-spin.
 
For the superconductor, 
we only require a 4-component Nambu basis $\ps^{\dagger}_{\bs R} = \left( \phi^{\dagger}_{\bs R \uparrow}, \phi^{\dagger}_{\bs R \downarrow}, \phi_{\bs R \downarrow}, -\phi_{\bs R \uparrow} \right)$.
In this basis, the Hamiltonian for superconductor, Eq~(\ref{eq:hs}), is -
\begin{align}
H_S & = \sum_{\bs R} \ps^{\dagger}_{\bs R} \left( \epsilon_{\text{sc}} \eta^z + \Delta \eta^x \right) \mathbf{I}_{\sigma} \ps_{\bs R} \non \\
 & \quad - t_{\text{sc}} \sum_{\langle \bs R, \bs R' \rangle} \ps^{\dagger}_{\bs R} \eta^z \mathbf{I}_{\sigma} \ps_{\bs R'} + h.c.
\end{align}
The  coupling between the semi-metal fermions and  the fermions in the superconductor, Eq~(\ref{eq:ht}), is -
\beq H_T = \sum_{\bs r, \bs R} \p_{\bs r}^{\dagger} A_{{\bs r},{\bs R}} \ps_{\bs R} + h.c. \eeq
where, 
\beq A_{{\bs r},{\bs R}} = \delta_{\bs r, \bs r_c} \delta_{\bs R, \bs r+\bs z} \left( 
    \begin{smallmatrix}
      \tilde{t}_1 & 0 & 0 & 0 \\
      0 & \tilde{t}_1 & 0 & 0 \\
      \tilde{t}_2 & 0 & 0 & 0 \\
      0 & \tilde{t}_2 & 0 & 0 \\
      0 & 0 & -\tilde{t}_1 & 0 \\
      0 & 0 & 0 & -\tilde{t}_1 \\
      0 & 0 & -\tilde{t}_2 & 0 \\
      0 & 0 & 0 & -\tilde{t}_2 
    \end{smallmatrix} \right)~. \eeq

\noindent
The total action for the system can now be written as 
\beq
S = \int_{-\infty}^{\infty} dt \left[ \sum_{{\bs R}} \ps_{\bs R}^{\dagger} \left[ i \hbar \partial_t \right] \ps_{{\bs R}} 
      + \sum_{{\bs r}} \p_{{ \br}}^{\dagger} \left[i \hbar \partial_t \right] \p_{{ \br}}  - H \right] \nonumber \eeq
 where $H$ is the complete Hamiltonian given in Eq.~(\ref{eq:Ah}). After taking a Fourier transform over time, the action can
 be written in terms of the bare Green's functions of the superconductor and the WSM as follows -
\bea
S &=& \int_{-\infty}^{\infty} \frac{d\omega}{2\pi} \sum_{\bs R, \bs R', \bs r,\bs r'} [ \ps_{\bs R}^{\dagger} \mathcal{G}^{-1}_{\bs R, \bs R'}(\omega) \ps_{\bs R'} + 
						\p_{\bs r}^{\dagger} G_{B \bs r,\bs r'}^{-1}(\omega) \p_{\bs r'} \non \\  
						& & +\quad  \ps_{\bs R}^{\dagger} A^{\dagger}_{\bs R,\bs r} \p_{\bs r} + \p_{\bs r}^{\dagger} A_{\bs r,\bs R} \ps_{\bs R} ] \eea
where, $\mathcal{G}(\omega)$ and $G_B(\omega)$ are the bare Green's functions of the superconductor and the semimetal respectively.
We now define 
\beq
\X_{\bs R} = \ps_{\bs R} + \sum_{\bs r, \bs R'} \mathcal{G}_{\bs R,\bs R'}(\omega) A^{\dagger}_{\bs R', \bs r} \p_{\bs r} \nonumber
\eeq
and obtain 
\beq
S = \int_{-\infty}^{\infty} \frac{d\omega}{2\pi} \left[ \sum_{\bs R, \bs R'} \X^{\dagger}_{\bs R} \mathcal{G}^{-1}_{\bs R, \bs R'} (\omega) \X_{\bs R'} + 
						\sum_{\bs r, \bs r'} \p_{\bs r}^{\dagger} G_{\bs r,\bs r'}^{-1}(\omega) \p_{\bs r'} \right]  \nonumber
						\eeq
where $G^{-1}(\omega) = G^{-1}_{B}(\omega) - \Sigma(\omega)$ is the inverse of the full Green's function and the self energy is 
given by
\bea
 \Sigma_{\bs r, \bs r'}(\omega) = \sum_{\bs R, \bs R'} A_{\bs r,\bs R} \mathcal{G}_{\bs R,\bs R'}(\omega) A_{\bs R', \br'}^{\dagger} \eea
The two fields $\X$ and $\p$ are decoupled and the effect of superconductor on the WSM  is encoded in the self-energy term. 
Now, we approximate $\mathcal{G}$ by the bulk Green's function of the superconductor given by
\begin{align}
\mathcal{G}_{\bs R,\bs R'}(\omega) & = \sum_{\bs K} \frac{e^{i \bs K \cdot (\bs R - \bs R')}}{\omega^2 - (\xi^2_K + \Delta^2)} \nonumber \\
& \quad \times \left(
		\begin{smallmatrix}
			\omega + \xi_K & 0 & \Delta & 0 \\
			0 & \omega + \xi_K & 0 & \Delta \\
			\Delta & 0 & \omega - \xi_K & 0 \\
			0 & \Delta & 0 & \omega - \xi_K
		\end{smallmatrix}  \right)~, \nonumber 
\end{align}
where $\bs K$ is the momentum in all three dimensions and $\xi_K = \epsilon_{\text{sc}} - 2t_{\text{sc}}\sum_{i} \cos{K_i}$. 
This approximation essentially ignores any surface effects that might exist in the superconductor itself. We ignore these, since the 
surface effects of a superconductor are not of primary interest here. 

Then assuming $\tilde{t}$ to be the same for both the orbitals, the on-site ($\bs r = \bs r'$) self-energy is
\bea
\Sigma_{\bs r, \bs r} (\omega) &=& (\tilde{t})^2 \delta_{\bs r, \bs r_c}  
\sum_{\bs K} \frac{1}{\omega^2 - (\xi^2_K + \Delta^2)} \times \nonumber \\
			&& \left[ \omega \mathbf{I}_{\eta} - \Delta \eta^x + \xi_K \eta^z \right] 
			  \left(\mathbf{I}_{\tau} + \tau^x \right)\mathbf{I}_{\sigma} \eea
Summing over the momenta, we get the final expression for self energy to be 
\beq
\Sigma_{\bs r} (\omega) = \delta_{\bs r, \bs r_c} \frac{\pi N(0) (\tilde{t})^2}{\sqrt{\Delta^2 - \omega^2}}	
\left[ \omega \mathbf{I}_{\eta} - \Delta \eta^x \right] \left(\mathbf{I}_{\tau} + \tau^x \right)
\mathbf{I}_{\sigma} \eeq
as used in the main text.


%
%

\end{document}